\documentclass[useAMS,usenatbib]{mn2e}
\usepackage{graphicx}
\usepackage{graphics}
\usepackage{epstopdf}
\usepackage{float}
\usepackage{bm}
\usepackage{epsf}
\usepackage{url}

\title[Locally Cold Flows from Large-Scale Structure]{Locally Cold Flows from Large-Scale Structure}
\author[Aragon-Calvo M.A. et al.]{M.A. Aragon-Calvo$^{1}$\thanks{E-mail:miguel@pha.jhu.edu} ,  J. Silk$^{2,1}$, A. S. Szalay$^{1}$\\
$^{1}$The Johns Hopkins University, 3400 Charles St., Baltimore, MD, USA.\\
$^{2}$Oxford University, Oxford, OX1 3RH.\\}
\begin{document}

\date{Submitted to MNRAS}

\pagerange{\pageref{firstpage}--\pageref{lastpage}} \pubyear{2002}
\maketitle
\label{firstpage}

\begin{abstract}
We show that the ``cold" Hubble flow observed for galaxies around the Milky Way does not represent a 
problem in cosmology but is due to the particular geometry and dynamics of our local wall.
The behavior of the \textit{perturbed} Hubble flow around the Milky Way is the result of two main factors: at small scales ($R < 1$ Mpc) the \textit{in}flow is
dominated by the gravitational influence of the Milky Way. At large scales ($R > 1$ Mpc) the \textit{out}flow reflects the expansion 
of our local wall which ``cools down" the peculiar velocities. This is an intrinsic property of walls and is independent of cosmology.
We find the dispersion of the local Hubble flow ($ 1 < R < 3 $ Mpc) around simulated ``Milky Way" haloes located at the centre of low-density 
cosmological walls to be $\sigma_H \sim $30 km s$^{-1}$, in excellent agreement with observations.
The expansion of our local wall is also reflected in the value of the measured \textit{local} Hubble constant. For ``Milky Way" haloes
inside walls, we find super-Hubble flows with $h_{\textrm{\tiny{local}}} \simeq 0.77-1.13$. 
The radius of equilibrium (R$_0$) depends not only on the mass of the central halo and the Hubble expansion 
but also on the dynamics given by the local LSS geometry. The 
super-Hubble flow inside our local wall has 
the effect of reducing the radius at which the local expansion balances the gravitational influence of the Milky Way. By ignoring 
the dynamical effect of the local wall, the mass of the Milky Way estimated from R$_0$ can be underestimated by as much as $\sim 30 \%$.
\end{abstract}

\section{Introduction}

The mystery of the ``Coldness of the Hubble Flow" (CHF) has remained largely unsolved despite many attempts to
explain or reproduce it. Early measurements of the radial velocities of nearby galaxies found the perturbed Hubble
flow dispersion to be $\sigma_H \sim 60$ km s$^{-1}$ \citep{Sandage72,Sandage86, Peebles88}.
Recent studies give an even lower value of $\sigma_H \sim 30$ km s$^{-1}$ \citep{Ekholm01,Karachentsev02,Karachentsev03,Karachentsev09}
This dispersion is significantly ``colder'' than theoretical expectations \citep{Peebles80, Scoccimarro04} and observed pairwise velocity dispersions which are
of the order of $300-500$ km s$^{-1}$ \citep{Zehavi02, Landy02, Hawkins03}.

\subsection{Searching for the origin of the CHF}

The scales involving the CHF suggest that it has a cosmological origin. Two main causes have been proposed to explain the CHF: 
background cosmology (in the form of a $\Lambda$ term) and  local environment. 
The influence of background cosmology was first studied in N-body simulations by \citet{Governato97} who failed to detect
low velocity dispersions in two models: $\Omega = 1$ (SCDM) and a $\Omega = 0.3$ (OCDM). A
later paper \citep{Baryshev01} 
proposed the role of the cosmological constant in the CHF under the assumption that the accelerating effect of $\Lambda$ or a dark energy term
results in a decay in the velocity fluctuations, thereby giving a colder local environment \citep{Baryshev01,Teerikorpi08,Peirani08, Maccio05, Chernin10}. 
On the other hand,  early constrained N-body simulations of the local group suggested that the observed low velocity dispersion 
somehow reflects our particular local environment  \citep{Weygaert00, Klypin03}. 
This possibility was also explored by \citet{Hoffman08} and \citet{Martinez09} who analyzed constrained simulations of the 
local group in several cosmologies and found the effect of the cosmological constant on local dynamics to be marginal.
While these studies reproduced to some degree cold Hubble flows they were not able to identify its
origin apart from being an intrinsic property of ``local-group" environments.

\section{Mimicking $\Lambda$ from local dynamics}
 
From the above discussion, a picture begins to appear: in order to cool the peculiar velocities, an accelerating $\Lambda$-like component
is needed, yet there are strong indications that this does not come from the background cosmology. 
This accelerating factor is somehow related to the particular local environment of the Milky Way.
One may ask under what circumstances one could have a local accelerated expansion without a $\Lambda$ component.
The most straightforward example is that of  cosmological voids. Their low density makes them
underdense island universes and their expansion adds to the drag coming from the global Hubble expansion 
as $H_{\textrm{\tiny{void}}} = 1.2 H_0$ \citep{Schaap07}.
This effect is independent of the background cosmology. It is purely the result of the
particular dynamics of underdense regions and, as we will see in the following section, it is not unique to voids.

\subsection{Local LSS geometry and dynamics} \label{sec:lss_dynamics}
In order to describe the role of the local geometry in the LSS dynamics it is illustrative to take a somewhat different approach to
the Zel'dovich collapse of an overdense region \citep{Zeldovich70} and focus on the draining and expansion of underdense valleys along
similar lines as \citet{Ike84, Ike87}. We start by generalizing the notion of a \textit{void} as an $n$-dimensional 
valley with $n=3,2,1$ corresponding to voids, walls and filaments respectively.  In this \textit{inverse} dynamical
description, matter is drained out of three-dimensional valleys (voids) and accumulates in 
two-dimensional (void-intersecting) walls. 
These two-dimensional valleys expand and lose their matter into their surrounding 
one-dimensional (wall-intersecting) filaments. 
Finally matter flows out of filaments into zero-dimensional (filament-intersecting) clusters.

The dynamics inside a wall can be roughly described as a two-dimensional expanding valley.
An observer situated at the centre of a wall will measure recession velocities $v_{rec}$ of surrounding 
galaxies as the sum of two components: $v_{rec} = v_{H}^{||} + v_{p}^{||}$,
where $v_{H}^{||}$ corresponds to the unperturbed Hubble flow and the line-of-sight peculiar velocity 
$v_{p}^{||} = v_{\textrm{\tiny{NL}}}^{||} + v_{\textrm{\tiny{LSS}}}^{||}$ 
is decomposed here into a non-linear component $v_{\textrm{\tiny{NL}}}^{||}$ coming from small-scale gravitational interactions
and a geometrical component $v_{\textrm{\tiny{LSS}}}^{||}$ originating from the particular dynamics of the local LSS geometry where the observer is located.
In the case of walls, the local expansion and draining of matter results in  $v_{\textrm{\tiny{LSS}}}^{||}$ being on average positive.
This extra component in the velocity is reflected in \textit{super-Hubble} flows which further cool down the local velocity fluctuations,
producing a \textit{cold} dynamical environment.

\begin{figure*}
  \centering
  \includegraphics[width=0.99\textwidth,angle=0.0]{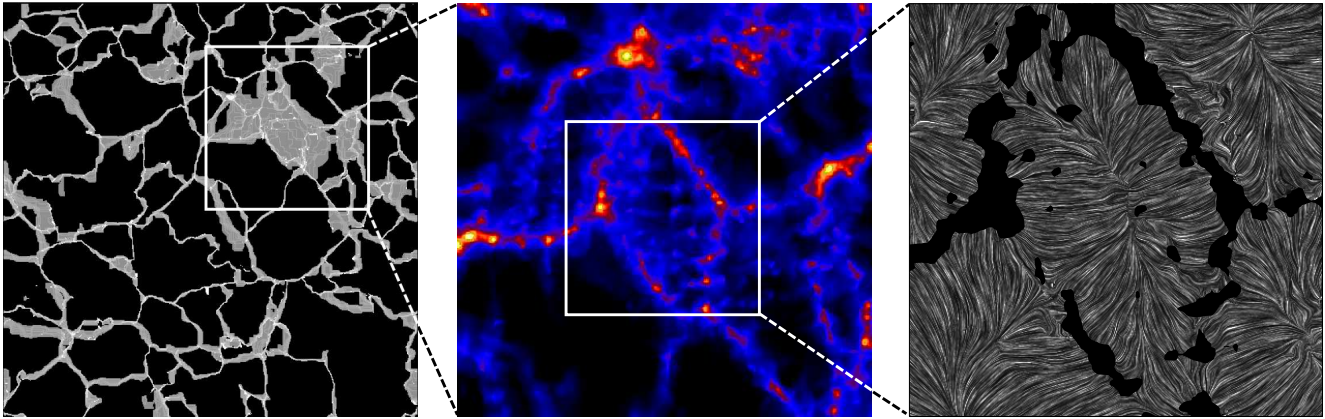}
  \caption{Local dynamics inside a cosmological wall. The three panels show different aspects of the geometry and dynamics of
  the Cosmic Web. Left panel: network of walls, filaments and voids from the SpineWeb method shown in a thick slice of 
  1.36 Mpc$/h$. 
  Central panel: Density field inside the
  zoomed region highlighted in the left figure. Right panel: local dynamics
  from the residual velocity field at scales of 8 Mpc$/h$ visualized with the LIC technique (see text for details). The lines represent streams 
  following the velocity field. The black filled areas correspond to filamentary regions with overdensity $\delta > 2$.}
  \label{fig:LIC}
\end{figure*}

\section{Are we living in a Wall?}

The Milky Way is the dominant member of the \textit{Local Group}, a system of galaxies forming a relatively flat pancake 
\citep{Hartwick00}. Extending perpendicular to the plane of the Milky Way, we find a planar association 
of galaxies up to $\sim$15 Mpc and at its edge the Virgo Cluster and its filamentary network \citep{Tully87}.
Perpendicular to the local wall, we find the local void extending up to $\sim40$ Mpc \citep{Tully08}. 
Our surrounding LSS geometry is consistent with the Milky Way being inside a wall of radius $r \sim 10$ Mpc.
It is interesting to note that the disk of the MW and its neighbours within 10 Mpc lie perpendicular 
to the plane of the local wall in excellent agreement with predictions from 
tidal torque theory \citep{Hoyle49,White84,Navarro04,Aragon07a}.

%
\section{N-body Simulation and DM haloes}

The analysis presented here is based on a dark matter N-body simulation with cosmological parameters 
$\Omega_m = 0.3, \Omega_{\Lambda} = 0.7, h = 0.73, \sigma_8 = 0.8$ and  200 Mpc$/h$ per side. 
Initial conditions were generated for $512^3, 256^3, 128^3$ and $64^3$ particles.
The simulation was run for all particle resolutions from $z=49$ until the present 
time, $z=0$ using the GADGET-2 N-body code \citep{Springel05}. From the final particle distribution of all resolutions
we compute the density field inside a cubic grid of 512 voxels per dimension using the 
DTFE method \citep{schaap00}.

We identify groups in the $512^3$ simulation at $z=0$ using the Friends of Friends group finder with a linking parameter $b=0.2$.
We impose a particle count limit of 100 corresponding to a halo mass of  $M > 5\times10^{11}$ M$_{\odot} h^{-1}$.
For each group, we compute its mass, radius, peculiar velocity, $v_{max}$, shape, etc. We then identify ``Milky Way" (MW)
haloes in the mass range $1\times 10^{12} $ M$_{\odot}$ $ <$ M$_{\textrm{\tiny{halo}}} < 3\times10^{12}$ M$_{\odot}$ and
maximum circular velocity $v_{\textrm{\tiny{max}}} > 20$ km s$^{-1}$.
 
%
\section{Finding walls in the Cosmic Web}
In order to isolate individual walls from the computed density field we performed the
hierarchical decomposition into voids, walls and filaments using the SpineWeb method \citep{Aragon10a}
extended to the hierarchical formalism described in \citet{Aragon10b}. 
In the hierarchical SpineWeb framework, each hierarchical level is defined by a physical smoothing scale in 
the linear regime and contains the full network of voids and its complementary web of walls and filaments. 
The hierarchical SpineWeb provides a complete parameter-free characterization of the 
geometry of the Cosmic Web. In the present analysis, we computed three hierarchical levels for practical reasons since further levels of
substructure tend to be noisy due to resolution effects. We generate catalogues of walls and filaments and for all 
pixels in walls we also compute the distance to its closest adjacent filament. 

Figure \ref{fig:LIC} illustrates the close relation between LSS geometry and local dynamics. The left panel shows the network
of structures found with the SpineWeb along a thick slice. A small region containing a face-on wall at its centre is highlighted by 
a white square and the corresponding density field is shown in the central panel. Note how the substructure inside the wall contrasts
with its adjacent empty voids. Even when the density of the wall is very low ($\delta \sim 0.5-1.0$) it is denser than the interior of its 
adjacent voids which appear completely dark in the central panel. The right panel shows the dynamics inside the central wall. 
We visualize the velocity field using the Line Integral Convolution (LIC) technique \citep{Cabral93}.
The velocity field shown here corresponds to the \textit{residual velocity} field $v_i^{\sigma}$  at a scale $\sigma$ defined as:
$v_i^{\sigma} = v_i - v_i \ast G(\sigma)$, where $v_i$ is the original velocity field decomposed in its three components and 
$v_i \ast G(\sigma)$ is its convolution with a Gaussian kernel of scale $\sigma=8$ Mpc$/h$. 
The velocity flows delineated by the LIC streams closely resemble the inverse dynamical description given in section 
\ref{sec:lss_dynamics}: matter flows out of walls into their adjacent filaments and from there to the clusters.
The outflow pattern at the interior of walls is remarkably coherent and offers a geometric origin for an extra expansion factor. 
This outflow is simply the result of the LSS geometry-dynamics as such it is independent of the background cosmology.

\begin{figure*}
  \centering
  \includegraphics[width=0.99\textwidth,angle=0.0]{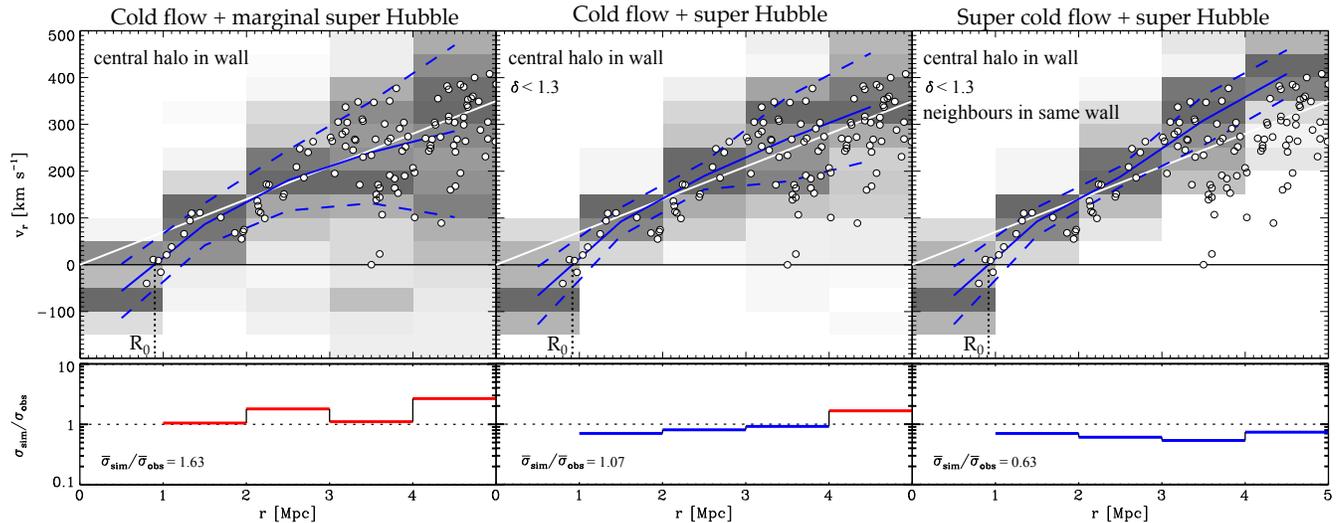}
  \caption{Cold and super-Huble flows inside walls. We show the distance-velocity diagram for central ``Milky Way'' haloes inside walls considering
  three increasingly ``cold" neighbourhoods:
    Top left panel shows the contribution from all haloes around central halos in walls. Top central panel is the same as left panel but considering only central halos with
    $\delta < 1.3$. Top right panel shows the case of $\delta < 1.3$ and including only the contribution from neighbour 
    haloes inside the \textit{same parent wall} as the centre halo. 
    The distribution of neighbour haloes is binned and presented 
    as grayscale in the background. Blue line corresponds to the mean in the corresponding bin computed 
    from a Gaussian fit.  Blue dashed lines indicate the standard deviation computed from the  Gaussian fit. 
    Filled circles correspond to galaxies given in \citet{Karachentsev09} for $r<3$ Mpc and  \citet{Karachentsev03} for $r>3$ Mpc. 
    The white solid line indicates a pure Hubble flow with $h=0.73$.
    The radius of zero velocity $R_0$ is also shown as vertical black dotted lines.
    Bottom panels show the corresponding ratios between the dispersion from N-body ($\sigma_{\textrm{\tiny{sim}}}$) and the observed one 
    ($\sigma_{\textrm{\tiny{obs}}}$). Values larger than one (hot) are plotted in red and blue (cold) otherwise. The blue color in the top panels
    indicates that for these curves $\bar{\sigma}_{\textrm{\tiny{sim}}} / \bar{\sigma}_{\textrm{\tiny{obs}}} < 2$ (see main text for details).}
  \label{fig:hubble_walls}
\end{figure*} 

%
\section{The Hubble Flow in Walls}

Having defined the halo catalog we proceed to select those halos inside walls and filaments. Central MW halos in walls were restricted to be
at least 3 Mpc away from their closest filament. No restriction was imposed on the neighbour haloes, except $v_{\textrm{\tiny{max}}} > 20$ km s$^{-1} h^{-1}$. 
We also computed the local density fluctuation ($\delta = \rho / \bar{\rho}$) around each halo from the dark matter density distribution using a Gaussian kernel 
with $\sigma_G = 3$ Mpc. We defined low ($\delta < 1.3$) and high ($\delta >2$) density environments 
roughly corresponding to voids-walls and filaments-clusters respectively (see Fig. \ref{fig:LIC}). 
For each MW halo we compute line-of-sight velocities as 
$v_{\textrm{\tiny{los}}} = \hat{\boldsymbol{r}} \; . \; \boldsymbol{v}_{\textrm{\tiny{MW-nei}}}$,
where $\hat{\boldsymbol{r}}$ is the unit vector between the central MW halo and its neighbour halo and 
$\boldsymbol{v}_{\textrm{\tiny{MW-nei}}}$ is the relative velocity between the two haloes. The radial velocity is then used to
compute the distance-velocity diagrams described in the following sections.

\subsection{Cold and super-cold Hubble flows in walls}

The Hubble flow for haloes in walls is shown in the distance-velocity diagrams of Fig. \ref{fig:hubble_walls} for three  increasingly cold and super-Hubble cases
(compare to Fig. \ref{fig:hubble_filaments}).
The flow for haloes in walls (left panel) has a mean velocity dispersion of $\sigma_H \simeq 55$ km s$^{-1}$  in the range $1 < r < 3$ Mpc after 
which point we observe a broadening in the dispersion up to $\sigma_H = 180$ km s$^{-1}$ at $r=5$ Mpc. 
The ratio between the simulated and
observed mean dispersions is $\bar{\sigma}_{\textrm{\tiny{sim}}} / \bar{\sigma}_{\textrm{\tiny{obs}}}  = 1.63$.
When the low-density condition is included (central panel) the Hubble flow becomes significantly colder with $\sigma_H \simeq 30$ km s$^{-1}$ at
$1 < r < 3$ Mpc. The broadening in $\sigma_H$ at larger distances is also smaller with $\sigma_H = 114$ km s$^{-1}$ at $r = 5$ Mpc
and the ratio between mean dispersions is $\bar{\sigma}_{\textrm{\tiny{sim}}} / \bar{\sigma}_{\textrm{\tiny{obs}}} = 1.07$.
The agreement of this particular sample and the observed galaxy population is remarkable with very similar dispersion across
the full range of the diagram and even showing the same broadening at $r \sim 3$ Mpc. 
Finally, if one further restricts the neighbour haloes to be located in the same parent wall as the central MW halo (right panel) we then have the extreme case of 
\textit{super-cold} flows with mean velocity dispersion $\sigma_H \simeq 25$ km s$^{-1}$ in the range $1 < r < 3$ Mpc. 
The low velocity dispersion is observed even at the boundaries of the largest walls in our sample with 
$\sigma_H = 50.9$ km s$^{-1}$ at $r = 5$ Mpc. The dispersion in this sample is significantly lower than the observed by almost a factor of two
($\bar{\sigma}_{\textrm{\tiny{sim}}} / \bar{\sigma}_{\textrm{\tiny{obs}}} = 0.63$).

\subsection{Super Hubble flows in walls}

The internal expansion of the walls is also observed in the distance velocity diagrams of Fig. \ref{fig:hubble_walls}.
The flow for haloes in all cases is sub-Hubble at small scales $r < 2$ Mpc reflecting the decelerating influence of the central halo.
For haloes in walls (left panel) the flow is marginally super-Hubble at $ r \sim 2.5$ Mpc.  
Once we include the low-density condition (central panel) the super-Hubble character of the
flow becomes more clear. At distances $r > 2$ Mpc we have a \textit{local} Hubble constant of $h_{\textit{\tiny{local}}} = 0.77$.
When we consider only neighbour haloes inside the same parent wall (right panel), then the effect of the local LSS geometry-dynamics 
dominates the distance-velocity diagram. The local Hubble constant is $h_{\textit{\tiny{local}}} = 1.13$  at scales $r > 2$ Mpc, 
which is $150\%$ larger than the global Hubble constant given by the background cosmology ($h=0.73$).
This local Hubble constant is also significantly larger than the expected for galaxies inside voids where 
$h_{\textrm{\tiny{void}}} = 1.2 h = 0.88$ \citep{Schaap07}.

\subsection{Radius of zero velocity ($R_0$)}

We explored the effect of the local geometry-dynamics on the radius at which no recession velocity is measured ($R_0$). This radius
depends to first order on the mass of the central halo as $R_0 \propto M^{1/3}$. The value of $R_0$ is basically independent of the local density
(see Fig. \ref{fig:hubble_filaments}, left and central panels) at $R_0 \simeq 1$ Mpc. 
However, when one divides the MW halo sample in terms of their surrounding LSS geometry, the effect of the local LSS dynamics emerges.
For MW halos in walls $R_0 \simeq 0.9$ Mpc (see Fig. \ref{fig:hubble_walls}). This reflects the fact that neighbour haloes recede faster 
than the global Hubble flow, thus making the central halo appear less massive.  The difference in mass from
no LSS-geometry distinction and haloes in walls is 
$M_{\textrm{\tiny{wall}}} / M_{\textrm{\tiny{all}}} \propto R_{0,\textrm{\tiny{wall}}}^3 / R_{0,\textrm{\tiny{all}}}^3 \simeq 1^3 / 0.9^3 = 0.73$.
Without considering the local geometry for central halos in walls, one can underestimate its mass by as much as $ \sim 27 \%$.

An extreme example of the effect of local geometry-dynamics on the value of $R_0$ is shown in the right panel of Fig. \ref{fig:hubble_filaments}
where we computed the distance-velocity diagram for central haloes located in filaments and high density regions and considering \textit{only}
neighbouring haloes located in walls. This case is the opposite dynamically to the one of haloes in walls.
The infall of haloes in filaments along the adjacent walls is reflected in the extremely high value of
$R_0$. This effect is independent of the mass of the central halo (compare to the central panel of Fig. \ref{fig:hubble_filaments}) and
if used to estimate the mass of the central MW halo can overestimate its mass by a factor of 
$M_\textrm{\tiny{est}} / M_\textrm{\tiny{real}} = 337 \%$ compared to the low and high density samples
(left and central panels of Fig. \ref{fig:hubble_filaments}) and 
$M_\textrm{\tiny{est}} / M_\textrm{\tiny{real}} = 462 \%$ compared to haloes in walls.

\subsection{Effect of density on the Hubble flow}

\begin{figure*}
  \centering
  \includegraphics[width=0.99\textwidth,angle=0.0]{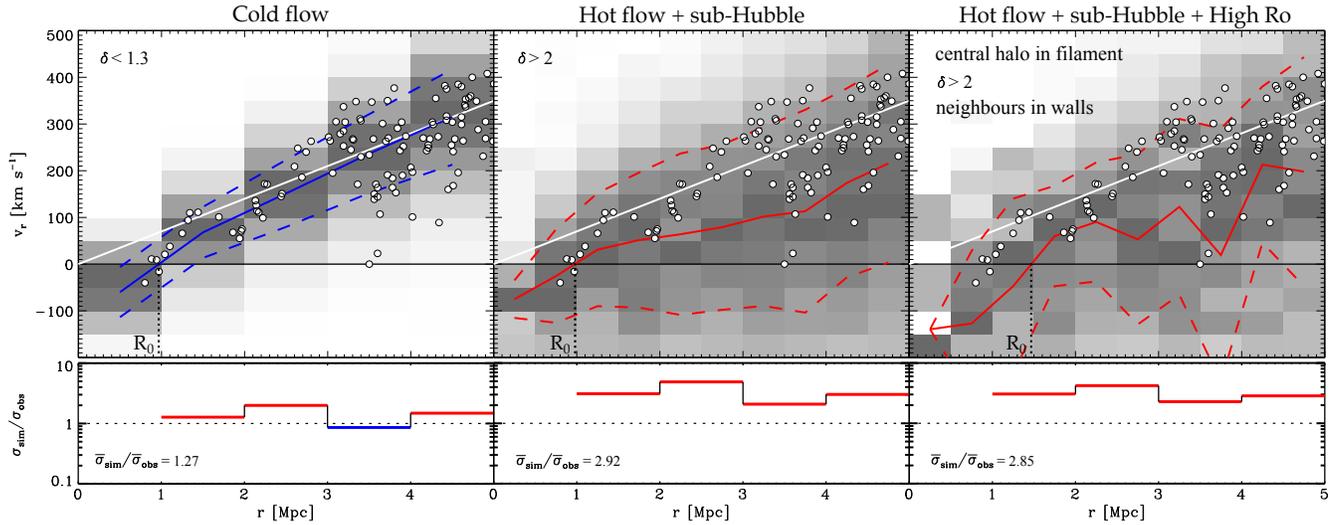}
  \caption{Distance-velocity diagram for MW haloes for three different environments: left panels:
  central haloes in low density regions. Central panels: central haloes in high density regions. Right panels: central haloes in
  high density regions and located inside filaments and considering only neighbour haloes located inside walls.
  The figure description is the same as in Fig. \ref{fig:hubble_filaments}.
   The blue color in the top-left panel indicates that for this curve $\bar{\sigma}_{\textrm{\tiny{sim}}} / \bar{\sigma}_{\textrm{\tiny{obs}}} < 2$.
   Top central and top right panels are colored in red indicating $\bar{\sigma}_{\textrm{\tiny{sim}}} / \bar{\sigma}_{\textrm{\tiny{obs}}} > 2$.}
  \label{fig:hubble_filaments}
\end{figure*}

Figure \ref{fig:hubble_filaments} shows the effect of local density on the distance-velocity diagram. Low density regions (left panel) have 
low velocity dispersions with $\sigma_H = 63.2$ km s$^{-1}$ in the range $1 <  r < 3$ Mpc. The value of $\sigma_H$ remains relatively
constant for all distances. The mean velocity converges to the global Hubble flow at $r \simeq 4$ Mpc. High density regions are
sub-Hubble and significantly hotter with $\sigma_H = 153.8$ km s$^{-1}$ going as high as $\sigma_H = 212.1$ km$/$s at $r=5$ Mpc.

\section{Conclusions and future work}
The coldness of the local Hubble flow is a longstanding problem in $\Lambda$CDM cosmology. We have shown that by taking account of 
the void/wall/filament nature of large-scale structure, one can naturally reproduce the observed low dispersion in the local Hubble flow
for galaxies located in local-group LSS configurations, corresponding to walls.
The dispersion of the local Hubble flow ($ 1 < R < 3 $ Mpc) around simulated ``Milky Way" haloes which are located at the centre of low-density 
cosmological walls is found  to be $\sigma_H \sim $30 km s$^{-1}$ (Fig. \ref{fig:hubble_walls}, central panel), in excellent agreement with observations.
The expansion inside walls is also reflected in the enhanced value of the measured \textit{local} Hubble constant. For ``Milky Way" haloes
inside walls, we find super-Hubble flows with $h_{\textrm{\tiny{local}}} \simeq 0.77-1.13$. Ignoring the dynamical effects of the local wall, 
the mass of the Milky Way estimated from R$_0$ can be underestimated by as much as $\sim 30 \%$.
In a future paper we will extend our study to include voids and filaments.

The strong constraint on the LSS dynamics imposed by the complementary LSS geometry can also be exploited to
improve current techniques of velocity field reconstruction. A full LSS morphological decomposition of the
local supercluster can allow us to obtain a more complete characterization of our local velocity field.

\section{Acknowledgements}
This research was funded by the Gordon and Betty Moore foundation.

\end{document}